\documentclass[aps,
pre,twocolumn,superscriptaddress,amsmath,amssymb,showpacs]{revtex4}
\usepackage[]{epsfig}

\newcommand{\bra}[1]{\left\langle #1\right|}
\newcommand{\ket}[1]{\left|#1\right\rangle}

\def\cal#1{\mathcal{#1}}

\def\u{\uparrow}
\def\d{\downarrow}
\def\uu{\ket{\uparrow \uparrow}}
\def\ud{\ket{\uparrow \downarrow}}
\def\du{\ket{\downarrow \uparrow}}
\def\dd{\ket{\downarrow \downarrow}}
\def\p{\ket{+}}
\def\m{\ket{-}}

\begin{document}

\title{Facilitated spin models in one dimension: a real-space
renormalization group study}

\author{Stephen Whitelam} 

\affiliation{Rudolf Peierls Centre For Theoretical Physics, University of
Oxford, 1 Keble Road, Oxford, OX1 3NP, UK} 

\author{Juan P. Garrahan}

\affiliation{School of Physics and Astronomy, University of
Nottingham, Nottingham, NG7 2RD, UK}

\date{\today}

\begin{abstract}
We use a real-space renormalization group (RSRG) to study the low
temperature dynamics of kinetically constrained Ising chains (KCICs).
We consider the cases of the Fredrickson-Andersen (FA) model, the East
model, and the partially asymmetric KCIC.  We show that the RSRG
allows one to obtain in a unified manner the dynamical properties of
these models near their zero-temperature critical points.  These
properties include the dynamic exponent, the growth of dynamical
lengthscales, and the behaviour of the excitation density near
criticality.  For the partially asymmetric chain the RG predicts a
crossover, on sufficiently large length and time scales, from
East-like to FA-like behaviour.  Our results agree with the known
results for KCICs obtained by other methods.
\end{abstract}


\maketitle
\section{Introduction}

Kinetically constrained models
(KCMs)~\cite{Fredrickson-Andersen,Palmer-et-al,Jackle-Eisinger,Kob-Andersen}
are systems in which certain trajectories between configurations are
suppressed~\cite{Whitelam-Garrahan}.  As a result, they display
interesting slow dynamical behaviour
\cite{Schulz-Trimper,Sollich-Evans,Crisanti-et-al,Chung-et-al,Toninelli-et-al}.
Simple KCMs, like the facilitated kinetic Ising model introduced by
Fredrickson and Andersen~\cite{Fredrickson-Andersen} (hereafter the FA
model) and J\"{a}ckle and Eisinger~\cite{Jackle-Eisinger} (hereafter
the East model), display the slow, cooperative relaxation
characteristic of supercooled liquids near the glass
transition~\cite{Garrahan-Chandler,Berthier-Garrahan}.  For a general
review of KCMs see~\cite{Ritort-Sollich}.

In this paper we show that a simple real-space renormalization group
(RSRG) scheme~\cite{Hooyberghs} yields the dynamical properties of
facilitated spin models in one dimension, or kinetically constrained
Ising chains (KCICs), near their zero temperature critical
point~\cite{Whitelam-et-al}.  This behaviour is known from previous
work~\cite{Ritort-Sollich}.  The RSRG scheme provides a unified
framework for treating these systems, allowing one to obtain critical
dynamic exponents, and to visualise the RG flows of scaling variables
related to temperature and spatial asymmetry.

We proceed as follows.  In Section~\ref{pseudo-spin} we define KCICs,
and show, following Siggia~\cite{Siggia}, that they can be written in
terms of ``interacting quantum spins''.  We discuss in
Section~\ref{sec:rd} how this formalism admits a simple physical
interpretation in terms of reaction-diffusion processes.  In
Sections~\ref{sec:rg}--\ref{sec:bcic} we use a RSRG scheme to extract
the zero-temperature critical behaviour of KCICs, in the FA model
(Sec.~\ref{sec:fa}) and East model (Sec.~\ref{sec:east}) limits,
and for the case of finite asymmetry~\cite{Buhot-Garrahan}
(Sec.~\ref{sec:bcic}).  We find that the dynamical exponents for the
FA and East models are respectively $z=2$ and $z= \left(T\ln 2
\right)^{-1}$, in agreement with existing
results~\cite{Ritort-Sollich}.  We show that length scales in the FA
model grow as $\xi \sim e^{1/T}$ near the critical point $T=0$, while
for the East model there is no characteristic length scale.  We also
quantify the crossover of the KCIC with large but finite asymmetry
from East-like to FA-like behaviour.  In Section~\ref{sec:conclu} we
state our conclusions.

\section{KCIC: a pseudospin formulation} \label{pseudo-spin} The KCIC is defined
as follows~\cite{Fredrickson-Andersen,Jackle-Eisinger}.  Consider a chain of
$N$ Ising spins $\sigma_i = \pm 1$, in one space dimension, with Hamiltonian
$H=\frac{1}{2}\sum_i \sigma_i$.  We will take $N$ even, assume periodic
boundary conditions, and restrict the dynamics to flips of single spins that
have at least one nearest neighbour in the up state. The transition rates
depend on whether the facilitating up-spin $\uparrow$ lies to the left or the
right of the flipping spin: \begin{equation} \uparrow \uparrow
\stackrel{b(1-c)}{\longrightarrow} \downarrow \uparrow, \, \, \downarrow
\uparrow \stackrel{bc}{\longrightarrow} \uparrow \uparrow, \, \, \uparrow
\uparrow \stackrel{\tilde{b}(1-c)}{\longrightarrow} \uparrow \downarrow, \,\,
\uparrow \downarrow \stackrel{\tilde{b} c}{\longrightarrow} \uparrow \uparrow,
\end{equation} where $b \in [0,1]$, $\tilde{b} \equiv 1-b$, and $c \equiv
(1+e^{1/T})^{-1} \approx e^{-1/T}$ at low temperature. The bias $b$ determines
the symmetry properties of the kinetic constraint: the FA and East models
correspond to the limiting cases of symmetry ($b=\frac{1}{2}$) and maximal
asymmetry ($b=0$), respectively. The East model is so-called because
information propagates to the east. We will also consider the case of general
$b$, which we will call the biased constrained Ising chain (BCIC).  For $b$
small but finite the BCIC exhibits a crossover at large length and time scales
from East-like to FA-like behaviour~\cite{Buhot-Garrahan}.  In
section~\ref{sec:rg} we use an RSRG to quantify this crossover. 

The dynamics of the KCIC is governed by the master equation
\begin{equation}
\label{one}
\frac{\partial P(\sigma,t)}{\partial t}=-\sum_i w(\sigma_i)
P(\sigma,t) + \sum_i w(-\sigma_i) P(\sigma',t),
\end{equation}
where $P(\sigma,t)$ is the probability that the system has
configuration $\sigma \equiv \{\sigma_1,
\dots,\sigma_i,\dots,\sigma_N\}$ at time $t$ ($\sigma'$ is the
configuration $\sigma$ with spin $\sigma_i$ flipped), and
$w(\sigma_i)\equiv w(\sigma_i,\{\sigma_j\})$ is the probability per
unit time that $\sigma_i$ will flip. The $\{\sigma_j\}$ are the
nearest neighbours of $i$. The matrix controlling the time development
of the $2^N$-component vector $P$ is not in general Hermitian, but can
be made so by introducing the vector $\psi(\sigma,t) \equiv
P_0(\sigma)^{\frac{1}{2}} P(\sigma,t)$. Here $P_0(\sigma)$ is the
equilibrium distribution. However, this obscures the fact that the
evolution operator is a normalized stochastic process which obeys
detailed balance, and so we will use the non-Hermitian representation
where this is explicit.

One passes to a quantum formalism~\cite{Siggia} by introducing the
state vector
\begin{equation}
\label{ket}
\ket{P(t)} = \sum_{\{\sigma^z\}} P(\sigma^z,t) \, \ket{\sigma_1^z}
\otimes \ket{\sigma_2^z} \otimes \cdots \otimes \ket{\sigma_N^z}.
\end{equation}
The ket $\ket{\sigma_i^z}$ is the state vector for the spin at site
$i$. The spins are for convenience taken to lie along the $z$-axis,
and thus the operator $\sigma_i^x$ flips the $z-$component of spin
$i$: $\sigma_i^x f(\sigma_i^z) = f(-\sigma_i^z) \sigma_i^x$. By
differentiating (\ref{ket}) with respect to time, and using
(\ref{one}) to eliminate $\dot{P}(\sigma,t)$, we get the master
equation in the guise of a Euclidean Schr\"{o}dinger equation,
\begin{equation}
\label{schrod}
\frac{\partial}{\partial t} \ket{P(t)}= -\mathcal{H} \ket{P(t)}.
\end{equation}
Here $\mathcal{H}$, which we will also call a ``Hamiltonian'', is not
in general Hermitian.

We define the KCIC by the constrained Glauber rates
\begin{equation}
w(\sigma_i,\{\sigma_{j}\})=\mathcal{C}_i(\{\sigma_{j}\})
\frac{e^{\beta \sigma_i/2}}{2 \cosh(\beta/2)},
\end{equation}
where the factor of $ 2 \cosh(\beta/2)$ is a convenient
normalization. The constraint is $\mathcal{C}_i(\{\sigma_{j}\})=
\tilde{b} n_{i-1} +b n_{i+1}$, where $n_i \equiv
\frac{1}{2}(1+\sigma_{i}^{z})$. The sigma matrices at a given site $i$
obey $\sigma_i^{\alpha} \sigma_i^{\beta} = \delta^{\alpha\beta} + i
\epsilon^{\alpha \beta}_{\gamma} \sigma_i^{\gamma}$.  Sigma matrices
at different sites commute. The matrix $\cal{H}$ in Equation
(\ref{schrod}) then reads
\begin{equation}
\label{matrixp}
\mathcal{H}= \cal{N} \sum_i \mathcal{C}_i(\{\sigma_{j}\})
\left(e^{\beta \sigma_i^z/2}-e^{-\beta \sigma_i^z/2}\sigma_i^x\right),
\end{equation}
where $\cal{N}^{-1} \equiv 2 \cosh (\beta/2)$. The appearance of the
$\sigma_i^x$ term shows that in order to represent these simple KCMs
in terms of ``interacting'' systems with no dynamical constraints, one
must introduce extra degrees of freedom, or, equivalently, nonlocal
interactions.  Using the Pauli representation
$\sigma^x=\left(\begin{array}{cc}0&1\\1&0 \end{array} \right)$ and
$\sigma^z=\left(\begin{array}{cc}1&0\\0&-1 \end{array} \right)$, we
have
\begin{equation} 
n_i \equiv \frac{1}{2}(1+\sigma_{i}^z)=\left( \begin{array}{cc} 1 & 0
     \\ 0& 0 \end{array} \right),
\end{equation} 
and
\begin{equation}  
\ell_i \equiv \frac{e^{\beta \sigma_i^z/2}-e^{-\beta \sigma_i^z/2}
     \sigma_i^x}{2 \cosh \beta/2}=\left( \begin{array}{cc} 1-c & -c \\
     c-1& c \end{array} \right).
\end{equation} 
The Hamiltonian $\mathcal{H}$ can then be written as the matrix direct
product
\begin{eqnarray}
\label{direct-product}
\mathcal{H}&=&(1-b)\sum_{i=1}^{N-1} \mathbf{1} \otimes \cdots \otimes
n_{i-1} \otimes \ell_i \otimes \mathbf{1} \otimes \cdots \otimes
\mathbf{1} \nonumber \\ &+&b\sum_{i=1}^{N-1} \mathbf{1} \otimes \cdots
\otimes \ell_{i-1} \otimes n_{i} \otimes \mathbf{1} \otimes \cdots
\otimes \mathbf{1} \nonumber \\ &\equiv& \sum_{i=1}^{N-1}
\mathcal{L}_i,
\end{eqnarray}
where the Liouvillian $\mathcal{L}_{i} \equiv \mathbf{1} \otimes
\cdots \otimes \mathcal{L} \otimes \cdots \otimes \mathbf{1}$ is
\begin{eqnarray} 
\label{interp}
\mathcal{L} &=& (1-b) n \otimes \ell + b \ell \otimes n \nonumber \\
&=&\left( \begin{array}{cccc} 1-c&-\tilde{b}c&-b c&0 \\
\tilde{b}(c-1)&\tilde{b}c&0&0\\ b(c-1)&0&bc&0\\ 0&0&0&0\end{array}
\right).
\end{eqnarray}
The matrix (\ref{interp}) describes a probability-conserving
stochastic process; thus the sum of each column is zero. When we
construct the RG scheme we must preserve this condition. The East and
FA models correspond to the cases $b=0$ and $b=\frac{1}{2}$,
respectively:
\begin{equation}
\label{liouv-east}
\mathcal{L}_E = n \otimes \ell = \left( \begin{array}{cccc}
     1-c & -c&0&0 \\
c-1&c&0&0\\
0&0&0&0\\
     0& 0&0&0  \end{array} \right),
\end{equation} 
and
\begin{equation}
\label{liouv-fa}
\mathcal{L}_{FA} =\frac{1}{2} \left(n \otimes \ell + \ell \otimes n
     \right)= \frac{1}{2}\left( \begin{array}{cccc} 2-2c & -c&-c&0 \\
     c-1&c&0&0\\ c-1&0&c&0\\ 0& 0&0&0 \end{array} \right).
\end{equation}

In the next section we show briefly that the evolution operators
obtained above have a simple physical interpretation as
reaction-diffusion processes.

\section{Interpretation via reaction-diffusion processes}
\label{sec:rd}

The interpretation of the KCIC as a reaction-diffusion process follows
by noting that the Liouville operators in the previous section act on
two-site basis states
\begin{equation}
\label{basis}
P(t)=\left( \begin{array}{c}
P(1) \\
P(0) \end{array} \right) \otimes \left( \begin{array}{c}
P(1) \\
P(0) \end{array} \right) =\left( \begin{array}{c}
P(1,1) \\
P(1,0)\\
P(0,1)\\
P(0,0) \end{array} \right),
\end{equation}
where $P(1)$ is the probability that a spin is up.  We have again
suppressed time labels.  The single-site basis states are normalized
probabilities of the form $P=\left( \begin{array}{c} \rho \\ 1-\rho
\end{array} \right)$, where $\rho <1$. The vectors $\left(
\begin{array}{c} 1 \\ 0 \end{array} \right)$ and $\left(
\begin{array}{c} 0 \\ 1 \end{array} \right)$ are the eigenvectors of
$\sigma^z$ with eigenvalue $+1$ and $-1$, respectively.

We represent reaction-diffusion processes as follows. Let the
eigenvalue $\sigma_i^z=1$ represent a lattice site $i$ occupied by a
particle $A$, and $\sigma_i^z=-1$ represent the same lattice site with
no particle present. We denote this state by $\emptyset$. Then the
East model Liouvillian (\ref{liouv-east}) has a clear physical
interpretation: it describes the (right) branching process
$A+\emptyset \rightarrow A+A$ occurring with rate $c$, and the (right)
coagulation process $A+A \rightarrow A+\emptyset$ occurring with rate
$1-c$. The FA model involves in addition the (left) branching and
coagulation processes $A+A \rightarrow \emptyset+A$ and $\emptyset +A
\rightarrow A+A$.

Consider the following general set of reaction-diffusion processes:
\begin{center}
 \begin{tabular}{|c|c|c|}
\hline
Process&Description&Rate\\
\hline
right diffusion & $A+\emptyset \rightarrow \emptyset+ A$ & $D_R$\\
left diffusion & $\emptyset+A \rightarrow A+\emptyset$ & $D_L$\\
right coagulation & $A+A \rightarrow A+\emptyset$ & $\gamma_{CR}$\\
left coagulation & $A+A \rightarrow \emptyset+ A$ & $\gamma_{CL}$\\
pair annihilation & $A+A \rightarrow \emptyset+ \emptyset$ & $\gamma_{A}$\\
right death & $\emptyset+A \rightarrow \emptyset+ \emptyset$ & $\delta$\\
left death & $A+\emptyset \rightarrow \emptyset+ \emptyset$ & $\delta$\\
right branching & $A+\emptyset \rightarrow A+A$ & $\rho_R$\\
left branching & $\emptyset+A \rightarrow A+A$ & $\rho_L$\\
pair creation & $\emptyset +\emptyset \rightarrow A+A$ & $\nu$\\
right birth & $\emptyset +\emptyset \rightarrow \emptyset+A$ & $\sigma$\\
left birth & $\emptyset +\emptyset \rightarrow A+\emptyset$ & $\sigma$\\
\hline
\end{tabular}
\end{center}
By inspection, using (\ref{basis}), the equation of motion for this system reads
\begin{equation}
\label{reac-diff}
\dot{P}(t)
= - \left( \begin{array}{cccc}
    \Gamma &-\rho_R&-\rho_L&-\nu\\
-\gamma_{CR}&\tilde{D}_R&-D_L&-\sigma\\
 -\gamma_{CL}&-D_R&\tilde{D}_L&-\sigma\\
-\gamma_A&-\delta&-\delta&\Sigma
 \end{array} \right) P(t),
\end{equation}
where $ \Gamma \equiv \gamma_{CR}+\gamma_{CL}+\gamma_{A}$,
$\tilde{D}_R \equiv D_R+\rho_R+\delta$, $\tilde{D}_L \equiv
D_L+\rho_L+\delta$ and $\Sigma \equiv 2\sigma+\nu$.

As an aside, we argued in Ref.~\cite{Whitelam-et-al} that a $d>1$
generalization of the FA model behaves like a system in the directed
percolation (DP)~\cite{Hinrichsen} universality class, albeit with
vanishing self-destruction rate. The latter process may be defined as
a reaction-diffusion system comprising diffusion, branching and
annihilation. The similarity between the FA model and DP is due to the
fact that nearest neighbour-facilitated branching can mimic the effect
of diffusion in higher dimensions, and thus the $d>1$ FA model can be
represented as (pseudo)-diffusion, branching and annihilation. In one
dimension this correspondence no longer holds: one can see from
Equation (\ref{reac-diff}) that left and right branching, $\rho_R$ and
$\rho_L$, and left and right diffusion, $D_L$ and $D_R$, sit in
different slots of the evolution matrix.

\section{Real-space renormalization group in $1+1$ dimensions}
\label{sec:rg}
We will now apply a simple real-space RG scheme to the KCIC. This
scheme was developed in the 1980s and used on quantum spin
models~\cite{Stella}. Recently, it was applied to the contact
process~\cite{Hooyberghs}, a reaction-diffusion system. Other RG
approaches that have been used on reaction-diffusion systems include
density matrix RG (DMRG)~\cite{Carlon-et-al} and field-theoretic RG
techniques. DMRG is a numerical scheme which tends to produce more
precise estimates for critical exponents than does the RSRG, but is
less intuitive, in the sense that it does not lend itself so readily
to the visualisation of flows in RG space. Field theoretic RG
techniques have been successfully applied to many reaction-diffusion
systems in low dimensions~\cite{Tauber,Cardy-rev}. However, we argued
in Ref.~\cite{Whitelam-et-al} that the FA model has an upper critical
dimension of 4, and, crucially, its coarse-grained action becomes
difficult to analyse below $d=2$. Hence we shall employ a real-space
RG scheme in $d=1$, which provides both an intuitive and a tractable
means of studying the critical behaviour of the KCIC.

The idea is as follows.  One divides the lattice into blocks of $p$
spins, and denotes the configuration of spins inside each block
$\alpha$ as $\{\sigma_i\}_{\alpha}$.  We will focus on the case $p=2$,
and discuss larger blockings where appropriate.  The evolution
operator $\cal{H}_P$ then splits into an intra-block part $\cal{H}_0$
and an inter-block `interaction' $V$:
\begin{equation}
\cal{H}=\sum_{\alpha} \left( \cal{H}_{0;\alpha} +V_{\alpha,\alpha+1} \right).
\end{equation}
In the case of the East model we can write
\begin{equation}
\label{intra}
\cal{H}_{0}=(n \otimes \ell) \otimes (1 \otimes 1 ),
\end{equation}
and
\begin{equation}
\label{inter}
V=(1 \otimes n) \otimes (\ell \otimes 1 ),
\end{equation}
where the brackets indicate the grouping of spins into blocks. All
terms in the Liouvillian (\ref{direct-product}) are of the form
(\ref{intra}) or (\ref{inter}), with the necessary number of identity
matrices affixed at each end of the chain.

We will denote the eigenstates of $\cal{H}_{0;\alpha}$ as $\ket{n}$ and
$\bra{n}$, noting that the left and right eigenstates of a non-Hermitian matrix
are in general different. The East and FA models have respectively triply- and
doubly-degenerate ground states, i.e. respectively three and two eigenvectors
with eigenvalue zero. One identifies these ground states as effective cell
states, and projects the basis of two-site spins $\ket{\sigma} \in
\{\uu,\ud,\du,\dd\}$ into renormalized block spins $\ket{\sigma'} \in
\{\p,\m\}$. This is done by defining a projection operator \begin{equation}
\hat{T}_1(\sigma', \sigma) \equiv \sum_{n,n'} c_{n, n'} \ket{n'} \bra{n},
\end{equation} where $\ket{n'}$ is a linear combination of the the renormalized
basis vectors $\ket{\sigma'}$, and the $c_{n,n'}$ are real numbers. If the
original Hamiltonian $\cal{H}$ is $2^N$ dimensional, then $T_1$ is a $2^{N/2}
\times 2^{N}$ matrix whose rows contain linear combinations of the left ground
states of $\cal{H}_0$. One can therefore write the matrix representation of
$\hat{T}_1$ as \begin{equation} \label{t1} T_1=\left(\begin{array}{cccc}
c_{\uu,\p}&c_{\ud,\p}&c_{\du,\p}&c_{\dd,\p}\\c_{\uu,\m}&c_{\ud,\m}&c_{\du,\m}&c_{\dd,\m} \end{array}\right)^{\otimes N/2}, \end{equation} where $c_{\uu,\p}$ is
the projection of the two-spin state $\uu$ onto the renormalized cell state
$\p$, etc. The projection operation is a mapping from an Hilbert space of $N$
spins, with dimension $2^N$, to an Hilbert space of $N/2$ block spins, with
dimension $2^{N/2}$. Thus $\hat{T}_1: \mathbb{C}^N \mapsto \mathbb{C}^{N/2}$. 
We also define the embedding operator \begin{equation} \hat{T}_2(\sigma,
\sigma') \equiv \sum_{n',n} c_{n',n} \ket{n}\bra{n'}, \end{equation} which may
be cast as a $2^{N} \times 2^{N/2}$ matrix whose columns are linear
combinations of the right ground states of $\cal{H}_0$. We have
\begin{equation} \label{t2} T_2=\left( \begin{array}{cc}
c_{\p,\uu}&c_{\m,\uu}\\c_{\p,\ud}&c_{\m,\ud}\\c_{\p,\du}&c_{\m,\du}\\c_{\p,\dd}&c_{\m,\dd} \end{array} \right)^{\otimes N/2}, \end{equation}  where now
$c_{\p,\uu}$ is the component of the renormalized cell state $\p$ that is
embedded in the two-spin state $\uu$. Thus $\hat{T}_2: \mathbb{C}^{N/2} \mapsto
\mathbb{C}^{N}$. We demand that if we apply the embedding operator followed by
the projection operator we recover the identity on the $N/2$-dimensional
block-spin space: $T_1 T_2 
=\mathbf{1}_{N/2}$. But since the projection operation does not retain
all the degrees of freedom of the system, projection followed by
embedding does not yield the identity on real-spin space. Thus $T_2
T_1 \neq \mathbf{1}_N$.

The renormalization prescription is then
\begin{equation}
\label{rg-prescription}
\cal{H}'(\sigma') = T_1(\sigma',\sigma)\, \cal{H}(\sigma)\, T_2(\sigma,\sigma'),
\end{equation}
where $\cal{H}'$ is a renormalized evolution operator. This prescription
projects the original $2^N \times 2^N$ Hamiltonian onto a $2^{N/2}
\times 2^{N/2}$ subspace. If this subspace is suitably chosen, the
renormalized Hamiltonian $\cal{H}'$ will have the same form as
$\cal{H}$, but with renormalized couplings $\lambda'=f(\lambda)$ and
rates $\Gamma'=p^z \Gamma$. From these relations one can determine
fixed points and critical exponents.

We will now apply this scheme to the KCICs.

\section{Renormalization of the FA model}
\label{sec:fa}

The FA model is defined by Equation (\ref{liouv-fa}).  When the
blocking parameter $p=2$, we need only consider the $16 \times 16$
matrix
\begin{eqnarray}
\label{fa-toy}
\cal{H}^{FA} &=& (n\otimes \ell) \otimes (1 \otimes 1)+(1\otimes n)
\otimes (\ell \otimes 1) \nonumber\\ &+& (\ell\otimes n) \otimes (1
\otimes 1)+(1\otimes \ell) \otimes (n \otimes 1).
\end{eqnarray}
The brackets indicate the groupings of cells into blocks. The first
and third terms in (\ref{fa-toy}) comprise the intra-cell Hamiltonian
$\cal{H}_0$; the second and fourth terms are the inter-cell
interaction $V_{\alpha,\alpha+1}$.

We must calculate the left and right ground states of the matrix
$\cal{H}_0$ (\ref{liouv-fa}). There are two left ground-state
eigenvectors, $(0,0,0,1)$ and $(1,1,1,0)$, and two right ground-state
eigenvectors, $(0,0,0,1)^T$ and $(c/(1-c),1,1,0)^T$. We will use these
to build $T_1$ and $T_2$, subject to the following constraints:
\begin{enumerate}
\item The RG transformation must preserve probability
conservation. Thus each column of $(\cal{H}^{FA})'$ must add up to
zero.
\item We require that $V'_{\alpha,\alpha+1}$ has the same form as
$\cal{H}_0$, so that we can identify unambiguously the renormalized
parameters. Note that by building $T_2$ from the ground states of
$\cal{H}_0$ we ensure that the renormalized intra-cell Hamiltonian
vanishes, i.e. $\cal{H}'_{0}=0$.
\item We must respect the fact that the FA model is {\em trivially
irreducible} for all $T \neq 0$~\cite{Ritort-Sollich}. This means that
any configuration (bar that with all spins down) can be reached from
an initial high-temperature configuration. This suggests that any
two-spin state with at least one up-spin, namely $\uu,\ud$ and $\du$,
should be projected onto $\p$.
\item Normalization. We require that $T_1
T_2=\left(\begin{array}{cc}1&0\\0&1 \end{array}\right)$.
\end{enumerate}
One choice satisfying these criteria is
\begin{equation}
\label{proj-fa}
T_1=\left(\begin{array}{cccc}
1&1&1&0\\
0&0&0&1 \end{array}\right); \quad
T_2=\frac{1}{2-c}\left(\begin{array}{cc}
 c&0\\
1-c&0\\
1-c&0\\
0&2-c\end{array}\right).
\end{equation}
The matrix $T_1$ projects $\uu,\ud$ and $\du$ onto $\p$, and $\dd$
onto $\m$. $T_2$ embeds the state $\p$ as $(2-c)^{-1} \left\{c
\uu+(1-c)\ud +(1-c) \du\right\}$, and $\m$ as $\dd$. The form of the
ground state vectors for the FA model stipulates that the states $\ud$
and $\du$ are treated on equal footing during the projection and
embedding operations, as befits a model whose dynamical rules are
isotropic. We will see in the next section that this is not so for the
East model.

\begin{figure}
\psfig{file=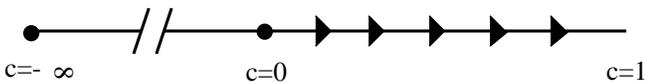,width=1.1cm,angle=270}
\caption{\label{fig:fa-flow} FA model RG flow diagram for the
temperature parameter $c = (1+e^{1/T})^{-1}$. The zero-temperature and
high temperature fixed points, $c^{\star}=0$ $(\lambda^{\star}=0)$ and
$(c^{\star}=1)$ $(\lambda^{\star}_a=\infty)$, are respectively
unstable and stable. The fixed point $c^{\star}=-\infty$
$(\lambda^{\star}=-1)$ is unphysical, and inaccessible.}
\end{figure}

Using (\ref{proj-fa}) and (\ref{liouv-fa}) we find 
\begin{equation}
\label{fa-R}
(\cal{H}^{FA})'=\frac{1}{2(2-c)}\left(\begin{array}{cccc}
2\frac{(1-c)^2}{2-c}&-c&-c&0\\
-\frac{(1-c)^2}{2-c}&c&0&0\\
-\frac{(1-c)^2}{2-c}&0&c&0\\
0&0&0&0\end{array}\right).
\end{equation}
We can deduce the flow of the temparature parameter as follows. Let
the ratio of the sum of the rates of the processes $A \emptyset
\rightarrow AA$ and $\emptyset A \rightarrow AA$ to the sum of the
rates of $ AA \rightarrow \emptyset A$ and $AA \rightarrow A
\emptyset$ be $\lambda$. Thus its unrenormalized or `bare' value is
\begin{equation}
\lambda_0 \equiv -\frac{(1,2)+(1,3)}{(1,1)} = \frac{c}{1-c} = e^{-1/T},
\end{equation}
where $(i,j)$ is element $(i,j)$ of matrix (\ref{liouv-fa}). Hence
$\lambda \rightarrow 0$ as $T \rightarrow 0$. We can work out how
$\lambda$ renormalizes by calculating a similar ratio using the matrix
(\ref{fa-R}). The resulting RG recursion relation is
\begin{equation}
\label{recursion-fa}
\lambda' = \lambda(2+\lambda),
\end{equation}
where $\lambda'$ is the renormalized counterpart of
$\lambda$. Equation (\ref{recursion-fa}) describes the flow of
$\lambda$ away from an unstable zero-temperature critical point
$\lambda^{\star}=0$, towards a stable high-temperature fixed point
$\lambda^{\star} \rightarrow \infty$. The unphysical fixed point
$\lambda^{\star}=-1$ is inaccessible. Figure~\ref{fig:fa-flow} shows this flow.

The RG procedure for the FA model using larger block sizes is
unambiguous, because one obtains at each stage only two right- and two
left ground state eigenvectors of the intra-block Hamiltonian. Thus
for $p=3$ we construct the following projection and embedding
operators:
\begin{equation}
\label{proj-fa-3}
T_1^{(3)}=\left(\begin{array}{cccccccc}
1&1&1&1&1&1&1&0\\
0&0&0&0&0&0&0&1 \end{array}\right),
\end{equation}
and
\begin{equation}
\label{embed-fa-3}
T_2^{(3)}=\cal{N}(\lambda)\left(\begin{array}{cc}
 \lambda^2&0\\
\lambda&0\\
\lambda&0\\
1&0\\
\lambda&0\\
1&0\\
1&0\\0&\cal{N}(\lambda)^{-1}\\
\end{array}\right),
\end{equation} 
where $\cal{N}(\lambda) \equiv \left(3+3
\lambda+\lambda^2\right)^{-1}$. In $T_2^{(3)}$, one inserts in the
relevant slot one power of $\lambda$ for every up-spin in excess of
one, in order to reflect the thermal suppression of these states. Thus
the state $\ket{\uparrow \uparrow \uparrow}$ (corresponding to element
$(1,1)$ of (\ref{embed-fa-3})) is penalized by a factor $\lambda^2$,
wheras the state $\ket{\downarrow \uparrow \uparrow}$ (corresponding
to element $(5,1)$ of (\ref{embed-fa-3})) receives a penalty of one
power of $\lambda$. The generalization to larger block sizes is
straightforward. We find that for general block size $p$ the RG
recursion relation is
\begin{equation}
\label{fa-rec-gen}
\lambda_k = (1+\lambda_{k-1})^p-1,
\end{equation}
where $\lambda_k$ is the value of $\lambda$ following the $k$-th
iteration of the RG. As expected (and required by the semi-group
property of the renormalization group) we see from
Equations~(\ref{recursion-fa}) and~(\ref{fa-rec-gen}) that two
successive coarse-grainings using a block size of $p=2$ is equivalent
to one coarse-graining using a block size of $p=4$. Thus $\lambda' = p
\lambda +\cal{O}(\lambda^2)$ near the critical point
$\lambda^{\star}=0$.

\begin{figure}
\psfig{file=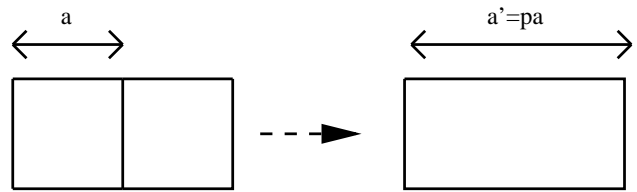,width=2.5cm,angle=270}
\caption{\label{fig:block} An illustration of how length scales change
under the RG blocking procedure. Say the system admits a dimensionful
correlation length $\xi_D$. Then in the unrenormalized system (left
panel) we may construct its dimensionless counterpart, measured in
terms of the lattice spacing $a$: $\xi=\xi_D/a$. In the renormalized
system (right panel) the new lattice parameter is $a'=pa$. Thus the
renormalized dimensionless correlation length is $\xi'=\xi_D/a'=p^{-1}
\xi$.}
\end{figure}

The divergence of the dynamical correlation length follows from
standard RG arguments~\cite{Cardy,Kadanoff}. Because the dimensionful
correlation length must remain invariant under the RG transformation,
the dimensionless correlation length $\xi$, measured in terms of the
lattice spacing, must decrease by a factor of the blocking parameter,
$p$: $\xi'=p^{-1} \xi$ (see Figure~\ref{fig:block}). We can write this
relation as
\begin{equation}
\label{rg-corr}
\xi(\lambda')=p^{-1} \xi(\lambda)
\end{equation}
In Equation~(\ref{rg-corr}) $\lambda'$ is the renormalized version of
$\lambda$. If we can write the RG equation for $\lambda$ near
criticality in the form $\lambda'=p^y \lambda +\cal{O}(\lambda^2)$,
then the correlation length is a function satisfying $\xi(p^y
\lambda)=p^{-1} \xi(\lambda)$. Hence $\xi(\lambda) \propto
\lambda^{-\nu_{\perp}}$, where $\nu_{\perp} \equiv 1/y$. From
Equation~(\ref{fa-rec-gen}) we see that $\nu_{\perp}=1$, and hence
near the critical point
\begin{equation}
\xi(\lambda_0) \sim \frac{1}{\lambda_0}=e^{1/T}.
\end{equation}

\begin{figure}
\psfig{file=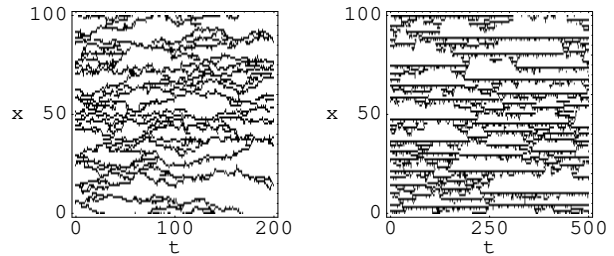,width=8.4cm,angle=0}
\caption{\label{fig:traj} Equilibrium space-time trajectories at
$T=1.0$ for the $1+1$-dimensional FA (left) and East (right) models,
reproduced from~\cite{Garrahan-Chandler}. Up-spins are black, down
spins white. Space runs along the vertical direction, encompassing
$10^5$ spins. Time runs along the horizontal. The characteristic
length scales of both systems correspond to the vertical extent of the
`bubbles' of down-spins, which scale in equilibrium as $l_{eq} \sim
c^{-1}$. The horizontal extent, $\tau$, of the bubbles is determined
by the relation $\tau \sim l^z$, where the dynamical exponents for the
FA and East models are $z=2$ and $z \propto 1/T$, respectively. As one
observes the trajectories shown above on smaller length and shorter
time scales, one moves from right to left along the RG flow diagrams
shown in Figures~\ref{fig:east-flow} and~\ref{fig:fa-flow}. Hence one
eventually probes behaviour controlled by the critical fixed
point. For the East model, the disappearance of the characteristic
length is consistent with the emergence of the fractal structure of
the bubble boundaries.}
\end{figure}

This correlation length corresponds to the characteristic spatial
extent of structures (`bubbles') in space-time trajectories of the FA
model at low temperature. We show one such trajectory in
Figure~\ref{fig:traj}.

We can obtain the dynamical exponent $z$ by noting that in the limit
of zero temperature the nonvanishing elements of (\ref{fa-R}) are
one-quarter those of (\ref{liouv-fa}). We find that for general $p$
the corresponding rescaling factor is $p^{-2}$. We interpret this
factor as a rescaling of time under renormalization, defining the
dynamical exponent $z$ via $t'= p^{-z} t$. Thus for the FA model
$z=2$, signifying diffusive behaviour. This is as expected: the
low-temperature dynamics of the FA model is known to proceed by
diffusion of isolated defects~\cite{Ritort-Sollich}.

We can infer the consequent relaxation time of the FA model by using
the relationship between time and length scales, $t \sim l^{z}$, where
$l$ is the length scale being probed.  Since the equilibrium length in
the FA model scales as $l_{eq} \sim c^{-1}$---see below, and
Refs.~\cite{Garrahan-Chandler,Ritort-Sollich}---and since the
microscopic timescale goes as $c$, we expect the equilibration time to
have the leading order temperature dependence $c \tau_{eq} \sim c^{-2}
\implies \tau_{eq} \sim \exp(3/T)$.  This scaling is known from
previous work on the FA model~\cite{Ritort-Sollich}.

One may also calculate~\cite{Hooyberghs} the density of excitated
sites, $n=\frac{1}{2} (1+\sigma)$, both in the steady state and near
the critical fixed point.  The former is trivial for the FA model,
since it obeys detailed balance, and one may therefore consider the
calculation of the steady-state density a test of the RG scheme.

First note that the renormalization of the number operator does not
depend on whether $n$ sits in the left or right slot of the block: $(1
\otimes n)'=(n \otimes 1)'=(2-c)^{-1} n_{\alpha}$, for a block size
$p=2$. The RG recursion relation for the density then reads
\begin{equation}
\label{fa-density}
n_k = \left(\frac{1+\lambda_k}{2+\lambda_k} \right) n_{k+1},
\end{equation}
where the subscript $k$ denotes the parameter obtained following $k$
iterations of the RG.

To extract the steady-state density we follow~\cite{Hooyberghs} and
write $n(\lambda_{k})=a(\lambda_k) n(\lambda_{k+1})$, where
$a(x)=(1+x)/(2+x)$. By iterating this equation along the RG flow we
get
\begin{equation}
\label{ss-density}
n_s(\lambda_0)=\left[ \prod_{i=0}^\infty a(\lambda_i) \right]
n(\lambda^{\star}_a),
\end{equation} 
where $n_s$ is the steady-state density, and $n(\lambda^{\star}_a)$ is
the density at the attractive fixed point $\lambda^{\star}_a =
\infty$. Again following~\cite{Hooyberghs}, we define $G_n(\lambda)
\equiv \prod_{i=0}^n a(\lambda_i)$. From
Equations~(\ref{recursion-fa}), (\ref{fa-density}) and the definition
of $a(\lambda)$ we can write
\begin{equation}
a(\lambda_k) = \frac{1}{2} \frac{d \ln \lambda_{k+1}}{ d \ln \lambda_k}.
\end{equation}
We can therefore write $G_n(\lambda)$ as
\begin{equation}
\label{gee}
G_n(\lambda)=\frac{1}{2^{n+1}} \frac{d \ln \lambda_{n+1}}{d \ln
\lambda_0}.
\end{equation}
From (\ref{recursion-fa}) we have that
$\lambda_{n+1}=(1+\lambda_0)^{2^{n+1}}-1$. Using this result with
Equations (\ref{ss-density}) and (\ref{gee}), we get
\begin{equation}
n_s(\lambda_0)=\lim_{n \rightarrow \infty}
\frac{\lambda_0}{\lambda_0+1} \left( 1+\frac{1}{\lambda_{n+1}} \right)
n(\lambda^{\star}_a).
\end{equation}
As $n \rightarrow \infty$, $\lambda_{n+1} \rightarrow \infty$, and so,
noting that $n(\lambda^{\star}_a)= 1$, we obtain the steady-state
density
\begin{equation}
n_s(\lambda_0)=\frac{\lambda_0}{\lambda_0+1} =c.
\end{equation}
This is as expected: detailed balance with respect to the Hamiltonian
$H(\sigma)=\frac{1}{2} \sum_j \sigma_j$ implies $\langle n_i
\rangle_{eq} = \sum_{\{\sigma\}} \left[\left(1+\sigma_i \right)/2
\right] e^{- \beta H(\sigma)}/ \sum_{\{\sigma\}} e^{-\beta H(\sigma)}
=c$.

Near criticality we can write $n(\lambda)=p^{-1}n(p \lambda)$, and so
$n(\lambda_0) \sim \lambda_0$. Thus the density vanishes close to
criticality as $n \sim c^{\beta}$ with $\beta = 1$.

\section{Renormalization of the East model}
\label{sec:east}

The East model is defined by Equation~(\ref{liouv-east}). To renormalize it
using a blocking parameter $p=2$, for example, we need only consider the $16
\times 16$ matrix \begin{equation} \label{east-toy} \cal{H}^E = (n\otimes \ell)
\otimes (1 \otimes 1)+(1\otimes n) \otimes (\ell \otimes 1). \end{equation} The
brackets indicate the groupings of cells into blocks. The first term in
(\ref{east-toy}) is the intra-cell component $\cal{H}_0$; the second is the
inter-cell interaction $V_{\alpha,\alpha+1}$. 

We must calculate the left and right ground states of the matrix
$\cal{H}_0$ (\ref{liouv-east}). The left ground states are represented
by the row vectors $(0,0,0,1),(0,0,1,0)$ and $(1,1,0,0)$. The right
ground states correspond to the column vectors $(0,0,0,1)^T,
(0,0,1,0)^T$ and $[c/(1-c),1,0,0]^T$. Next, we choose the projection
and embedding matrices, which we call $R_1$ and $R_2$ so as not to
confuse them with their FA model counterparts. One choice satisfying
criteria 1--4 (see above) is
\begin{equation}
\label{proj}
R_1=\left(\begin{array}{cccc}
1&1&1&0\\
0&0&0&1 \end{array}\right), \quad
R_2=\left(\begin{array}{cc}
a c&0\\
a(1-c)&0\\
1-a&0\\
0&1\end{array}\right),
\end{equation}
where $0<a<1$ parameterizes a degree of freedom. This arises because
the East model admits one more ground state vector than the FA
model. $R_1$ projects $\uu,\ud$ and $\du$ onto $\p$, and $\dd$ onto
$\m$. $R_2$ embeds the state $\p$ as $ac \uu+a(1-c)\ud +(1-a) \du$,
and $\m$ as $\dd$. With this choice we get
\begin{equation}
\label{east-R}
(\cal{H}^E)'=(1+ac-a)\left(\begin{array}{cccc}
a(1-c)^2&-c&0&0\\
-a(1-c)^2&c&0&0\\
0&0&0&0\\
0&0&0&0
\end{array}\right).
\end{equation}
We deduce the flow of the temperature parameter in a similar way to
before: let the ratio of the rates of the processes $A \emptyset
\rightarrow AA$ and $AA \rightarrow A \emptyset$ be $\lambda$. Then
\begin{equation}
\lambda_0 \equiv -\frac{(1,2)}{(1,1)} = \frac{c}{1-c} = e^{-1/T}
\end{equation}
where $(i,j)$ is element $(i,j)$ of matrix (\ref{liouv-east}). Hence
the bare temperature parameter has the same interpretation as in the
FA model. We can work out how $\lambda$ renormalizes by calculating
the ratio of elements $(1,2)$ and $(1,1)$ of matrix
(\ref{east-R}). The resulting RG recursion relation is
\begin{equation}
\label{recursion-east-0}
\lambda'=a^{-1} \lambda(1+\lambda),
\end{equation}
implying an unstable zero-temperature critical point,
$\lambda^{\star}=0$, as expected.

The dynamical exponent $z$ follows immediately. In the critical limit
$\lambda \rightarrow 0$, element $(1,1)$ of matrix (\ref{liouv-east})
becomes unity. Hence we may interpret the renormalized value of this
element as the time rescaling factor $2^{-z}$. From (\ref{east-R}) we
get
\begin{equation}
\label{zed}
2^{-z} = \lim_{c \rightarrow 0} \Big\{1+a(c-1) \Big\}a, 
\end{equation}
and so $z$ depends on the value we choose for $a$.

Let us choose $a$. This parameter measures the extent to which we
treat the states $\du$ and $\ud$ on equal footing. In a model with
symmetric dynamical rules, such as the FA model, we must treat these
states identically. But the East model has asymmetric dynamical rules,
suggesting that at some point in our calculation we must suppress
$\du$ relative to $\ud$, or vice-versa. At which point should we do
this? We note that the {\em projection} matrix $R_1$ treats $\ud$ and
$\du$ identically. If this were not the case, and we instead (for
example) used
\begin{equation}
R'_1=\left(\begin{array}{cccc}
1&1&0&0\\
0&0&1&1 \end{array}\right),
\end{equation}
we would violate criteria 2 and 3 above. ($R'_1$ imposes a symmetry
between flipping spins $\u \leftrightarrow \d$ in a two-spin block and
flipping the resulting renormalized spin $\p \leftrightarrow \m$.)
Therefore, we conclude that the {\em embedding} matrix $T_2$ must
treat $\ud$ and $\du$ asymmetrically. The simplest way of doing this
is to set $a=1$, thus suppressing completely the state $\du$. This
corresponds to the assertion that a spin configuration $\du$ (which is
unable to change state unless connected to neighbouring spins) is much
less important to the dynamics than a configuration $\ud$, which is
mobile. Thus when one renormalizes the lattice using $R_1$ and $R_2$
with $a=1$, one effectively discards dynamical pathways mediated by
blocks of `jammed' spins $\du$. The RG process discards inaccessible
pathways in trajectory space $\{\sigma_{t_1},\sigma_{t_2},\dots\}$,
according to rules imposed by the Liouvillian of the dynamical
process. Loosely, the projection matrix $R_1$ identifies those
single-spin states which are facilitat{\em ing}, whereas $R_2$ picks
out those two-spin states which are (internally) mobile.

Setting $a=1$ immediately yields a temperature dependent dynamic
exponent: from Equation~(\ref{zed}) we obtain $2^{-z}=c$, or $z=(T
\ln2)^{-1}$. Were $a<1$, $z$ would be independent of temperature to
leading order. We thus conclude that maximal spatial anisotropy in the
embedding process is a necessary condition for a temperature-dependent
dynamic exponent.

The RG scheme for the East model can be generalized to larger block
sizes. However, this procedure is less straightforward than for the FA
model, because of the freedom one is afforded by the East model's many
ground state eigenvectors. Furthermore, the results one obtains
depends on whether one coarse-grains using a blocking parameter $p$
equal to a power of $2$, or not.

Let us first illustrate the generalization of the procedure for the
case $p=4=2^2$. We show that the results are consistent with the $p=2$
scheme. We then argue that one should obtain a different dynamical
exponent if one coarse-grains the system using a block size not equal
to a power of $2$, and then show explicitly for $p=3$ that this is
indeed the case.

Consider $p=2^n$, where $n$ is an integer. Building $R_1^{(p)}$ is
straightforward: it is identical to $T_1^{(p)}$, its FA model
counterpart. Thus $R_1^{(4)}$ is a $2 \times 16$ matrix whose top row
is composed of $1$s apart from the rightmost element which is
zero. Vice versa for the bottom row.

The form of the embedding matrix is less obvious, because the number
of ground states increases as one increases the block size. However,
we are guided by the form of the Liouvillian, which for block size
$p=4$ may be written schematically as
\begin{eqnarray}
\label{east-l-4}
\cal{H}^{E} &=& (n\otimes \ell \otimes 1 \otimes 1)+(1\otimes 1
\otimes n \otimes \ell) \nonumber\\ 
&+& \qquad \qquad (1\otimes n \otimes \ell \otimes 1)  \\ 
&+& (\ell\otimes 1 \otimes 1 \otimes 1)+(1\otimes 1
\otimes 1 \otimes n). \nonumber
\end{eqnarray}
Brackets again denote the grouping of cells into blocks. We take the
first line of Equation~(\ref{east-l-4}) as the intra-cell Hamiltonian
$\cal{H}_0$. The second line vanishes under renormalization as a
consequence of $T_1$ acting on it from the left, and so we ignore it;
the third line comprises the inter-cell interaction whose
renormalization properties we wish to study.

Guided by the form of $R_2^{(2)}$, we find that one choice of $R_2^{(4)}$
satisfying criteria 1--4 above is \begin{equation} \label{embed-east-4}
R_2^{(4)}=\cal{N}^{(4)}(\lambda)\left(\begin{array}{cc}  \lambda^3&0\\
\lambda^2&0\\ 0& 0\\ \lambda&0\\ \lambda^2& 0\\ \lambda&0\\ 0&0\\ 1&0\\ 0&0\\
0&0\\ 0&0\\ 0&0\\ 0&0\\ 0&0\\ 0&0\\ 0&\cal{N}^{(4)}(\lambda)^{-1}
\end{array}\right), \end{equation} where $\cal{N}^{(4)}= \left(1+2 \lambda+2
\lambda^2+\lambda^3 \right)^{-1}$. We see that Equation~(\ref{embed-east-4})
can be obtained from its FA model counterpart by using a simple rule-of-thumb:
suppress all states of the form $\ket{\downarrow \cdots}$ [corresponding to
elements $(9,1)$--$(18,1)$ of (\ref{embed-east-4})], as well as states
possessing a `frozen' up-spin at the right-hand boundary of the block. Thus
states $\ket{\uparrow \uparrow \downarrow \uparrow}$ [corresponding to element
$(3,1)$ of~(\ref{embed-east-4})] and $\ket{\uparrow \downarrow \downarrow
\uparrow}$ [element $(7,1)$] have been removed. We see again that the embedding
operator plays the role of a dynamical `filter', eliminating those states which
play a sub-dominant role in the dynamics of the East model. 

With these choices of embedding and projection operators we obtain the RG
recursion relation for the temperature, \begin{equation} \label{east-rec-4}
\lambda_{(4)}'=\lambda \left(1+2\lambda+2 \lambda^2 +\lambda^3\right),
\end{equation} and a relation for the dynamical exponent: \begin{equation}
\label{zed-east-4} 4^{-z} = \lim_{\lambda \to 0} \frac{\lambda^2(1+
\lambda)}{\left(1+\lambda+\lambda^2 \right)^2}. \end{equation} Equation
(\ref{east-rec-4}) is identical to the result one would obtain via two
coarse-grainings using a block size $p=2$ [Equation (\ref{recursion-east-0})],
as required. Equation (\ref{zed-east-4}) yields the dynamical exponent
$z=\left(T \ln 2\right)^{-1}$, as before. 

We shall demonstrate how one can generalize this approach to
arbitrarily large $n$. Let us use reaction-diffusion notation $(\u \to
1,\d \to 0$), and write the projection and embedding operators in the
form
\begin{equation}
\hat{R}_1 = \p \left(\sum_{1\star} \bra{1\star} \right) + \m
\bra{00\cdots 0},
\end{equation}
and
\begin{equation}
\hat{R}_2 = \left(\sum_{1\star} a_{1\star} \ket{1\star} \right)
\bra{+}+ \ket{00\cdots 0} \bra{-}.
\end{equation}
The symbol $1\star$ denotes any state $\ket{1 \cdots }$ {\em starting}
with a $1$, and the $\{a_{1\star}\}$ are a set of coefficients [see
e.g. the first column of (\ref{embed-east-4})]; the projection
operator $\hat{R}_1$ allows one to compare this notation to the matrix
representations employed previously. The normalization requirement
$\hat{R}_1 \cdot \hat{R}_2 = \mathbf{1}$ implies $\sum_{1\star}
a_{1\star} =1$. Thus at least one of the coefficients $a$ must be of
$\cal{O}(1)$.

The values of these coefficients are fixed by the eigenvectors of the
intra-block evolution operator, as we have discussed. We can see how
these coefficients determine the properties of the model under
renormalization, as follows. We find that `bulk' states of the form
$(\cdots 1 \otimes \hat{n} \otimes \hat{\ell} \otimes 1 \cdots)$
vanish under renormalization as a consequence of the projection
operator acting from the left. We are therefore left with the
`surface' terms $(\hat{\ell} \otimes 1 \otimes \cdots \otimes 1)$ and
$(1 \otimes \cdots \otimes \hat{n})$, in which the operators $\hat{n}$
and $\hat{\ell}$ sit at the edge of the block. We find that, under
renormalization,
\begin{equation}
\label{haile}
\hat{R}_1 \cdot \left(1 \otimes \cdots \otimes \hat{n}\right) \cdot
\hat{R}_2 \to \left( \sum_{1\star1}
a_{1\star1} \right) \hat{n}'
\end{equation} 
and
\begin{equation}
\label{gebrselassie}
\hat{R}_1 \cdot \left(\hat{\ell} \otimes 1 \otimes \cdots \otimes
1\right) \cdot \hat{R}_2 \to \left(a_{10\cdots 0}\right) \,
\hat{\ell}'\left[ \lambda \to \frac{\lambda}{a_{10\cdots 0}}\right].
\end{equation} 
In Equations~(\ref{haile}) and~(\ref{gebrselassie}) primes denote
renormalized operators. The symbol $1\star1$ denotes
states $\ket{1 \cdots 1}$ starting {\em and ending} with a $1$. In
(\ref{gebrselassie}) the temperature parameter $\lambda$ has been
rescaled by the coefficient $a_{10\cdots 0}$ which weights thermally
the state $\ket{10 \cdots 0}$ with a single $1$ at the leftmost edge,
followed by a string of $0$s. From the previous discussion we know
that this coefficient is of order unity, and hence the recursion
relation for $\lambda$ will be marginal, as we have found.

The dynamical exponent follows by noting that the product of
Equations~(\ref{haile}) and~(\ref{gebrselassie}) constitutes the renormalized
evolution operator, and so the prefactor describes the rescaling of time as a
consequence of rescaling space. Thus \begin{equation} \label{komen} p^{-z}
\propto\lim_{\lambda \to 0} \left(a_{10\cdots0}\right) \times \left(
\sum_{1\star1} a_{1\star1} \right). \end{equation} The constant of
proportionality in (\ref{komen}) is $(1-c)^{-1}= 1+\lambda$, i.e. the
reciprocal of element $(1,1)$ in the unrenormalized East model
Hamiltonian~(\ref{liouv-east}). The first factor on the right hand side of
(\ref{komen}) is of order unity. The second factor is fixed by the embedding
operator, which is in turn determined by the relevant East model eigenvectors.
The rule-of-thumb we obtained above tells us that we remove from this factor
any state with a frozen rightmost up-spin. This may be regarded as an entropic
suppression of states playing only a sub-dominant role in the dynamics. Those
states starting and ending with a 1 which {\em are} important for the dynamics
of the East model are for block sizes $p=2,4,8$ and $16$, \begin{eqnarray} &&11
\left(\lambda \right), 1011 \left(\lambda^2 \right), 10001011 \left(\lambda^3
\right), \nonumber \\ &&1000000010001011 \left(\lambda^4 \right).
\end{eqnarray} All have a `mobile' rightmost up-spin. The thermal weighting of
each state is given in brackets. These states are important because of the
hierarchical dynamics of the East model~\cite{Ritort-Sollich,Sollich-Evans},
which dictates that two defects separated by a distance $d$ are relaxed by
establishing a set of isolated defects between them, at distances $d/2$, $3
d/4$ etc. Thus for block size $p=4$ the dominant dynamical pathway proceeds via
the state $1011$, with a thermal weighting of $\lambda^2$ (and not, for
example, $1001$, which has a weighting of $\lambda$). Hence $a_{1011} \sim
\lambda^2$, $a_{1001}=0$, and so $\lim_{\lambda \to 0} \sum_{1\star1}
a_{1\star1} \sim \lambda^2$. Consequently, the rescaling factor is $4^{-z} \sim
\lambda^2$, and the dynamical exponent $z=(T \ln 2)^{-1}$, as required. (In the
FA model, states such as $\ket{0001}$ are permitted, leading to
temperature-independent $a$ coefficients and hence to a temperature-independent
$z$.) Thus $R_2$, which attempts to reconstitute an unrenormalized state from a
coarse-grained state, captures both energetic effects (the powers of $\lambda$
weighting thermally the various states) and entropic effects (the `zero'
entries corresponding to those suppressed entropically). We conclude that the
RG scheme for the East model generalizes readily to larger block sizes.

It is interesting to note that if one uses blocks of size not equal to
a power of $2$, one obtains a slightly different result for the
dynamical exponent. We argue that this is a consequence of the
hierarchical dynamics of the East model taking place naturally in
blocks of lengths equal to a power of
$2$~\cite{Ritort-Sollich,Sollich-Evans}. We can derive the approximate
value of $z$ that one should obtain from a coarse-graining over block
sizes $p \neq 2^n$. Let us take $p=3$ as an illustration. Consider the
coarse-grained relaxation process $\ket{++}
\stackrel{\gamma'}{\longrightarrow} \ket{+-}$. We wish to determine
the leading order temperature dependence of $\gamma'$, noting that the
rate for the equivalent unrenormalized process, $\uu
\stackrel{\gamma_0}{\longrightarrow} \ud$, is $\gamma_0 =
\cal{O}(1)$. From our previous discussion of the form of the embedding
matrices, we can infer that the dominant dynamical pathway (involving
`unrenormalized' spins) contributing to this renormalized process is
$\ket{\uparrow \downarrow \downarrow \uparrow \downarrow \downarrow}
\stackrel{\gamma_1}{\longrightarrow} \ket{\uparrow \downarrow
\downarrow \downarrow \downarrow \downarrow}$. To relax the second
up-spin, one must create two extra up-spins to the right of the first
up-spin. Hence this pathway has a rate $\gamma_1 \sim c^2$, and the
renormalized rate $\gamma' = \cal{O}(c^2)$. Other pathways also
contribute to the renormalized process $\ket{++}
\stackrel{\gamma'}{\longrightarrow} \ket{+-}$, but do so either with
rates $\sim c^2$ (e.g. $\ket{\uparrow \uparrow \downarrow \uparrow
\downarrow \downarrow} \longrightarrow \ket{\uparrow \downarrow
\downarrow \downarrow \downarrow \downarrow}$)---in which case
$\gamma'$ is changed only by a temperature-independent numerical
factor---or with rates higher order in $c$ (e.g. the pathway
$\ket{\uparrow \uparrow \uparrow \uparrow \downarrow \uparrow}
\longrightarrow \ket{\uparrow \downarrow \uparrow \downarrow
\downarrow \downarrow}$). These we may ignore. Since we interpret the
overall rescaling of the fundamental relaxation rate deriving from a
coarse-graining of space as the numerical factor $p^{-z}$, we would
therefore expect for $p=3$ that $3^{-z} \sim c^2$, or $z \sim 2/(T \ln
3)$.

Loosely, then, we expect that by coarse-graining space in blocks of
size $2^{n-1}<p<2^{n}$ one should obtain $z \approx n/\left(T \ln p
\right)$ [which tends to $z \to (T \ln2)^{-1}$ when $p \to
\infty$]. Coarse-graining using block sizes $p=2^n$ yields $z=(T \ln
2)^{-1}$. This is as we expect: the energetic barriers for relaxing
chains of lengths $2^{n-1}<p<2^n$ and $p=2^{n}$ are identical, but the
{\em entropic} barriers are larger for the latter case. Thus one would
expect the dynamical exponents to differ. More sophisticated
arguments~\cite{Aldous-Diaconis} reveal that $z$ is bounded by $(T
\ln2)^{-1}$ and $(2 T \ln 2)^{-1}$.

We can show that our guess for the dynamical exponent is borne out in the case
$p=3$ by the RG scheme. We construct the embedding operator according to the
\begin{equation} \label{embed-east-3}
R_2^{(3)}=\cal{N}^{(3)}(\lambda)\left(\begin{array}{cc}  \lambda^2&0\\
\lambda&0\\ 0&0\\ 1&0\\ 0&0\\ 0&0\\ 0&0\\ 0&\cal{N}^{(3)}(\lambda)^{-1}
\end{array}\right), \end{equation}  where
$\cal{N}^{(3)}=\left(1+\lambda+\lambda^2 \right)^{-1}$. Together with the
obvious choice for the projection operator we find \begin{equation}
3^{-z}=\lim_{\lambda \to 0} \frac{\lambda^2}{\left(1+\lambda+\lambda^2
\right)^2}, \end{equation} yielding $z =2/\left(T \ln3 \right)$, as advertised.
We conclude that the RG scheme for the East model can be generalized to larger
block sizes, but more naturally so for the case of a blocking parameter $p$
equal to a power of $2$. For simplicity we shall focus on the case $p=2$. 

\begin{figure}
\psfig{file=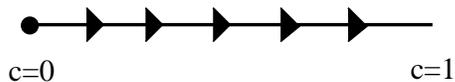,width=1.1cm,angle=270}
\caption{\label{fig:east-flow} East model RG flow diagram for the
temperature paramter $c = (1+e^{1/T})^{-1}$. The zero-temperature
critical fixed point $c^{\star}=0$ $(\lambda^{\star}=0)$ is
unstable. Near this fixed point, corresponding to observations of the
system on small length and time scales, the East model admits no
characteristic length. This is consistent with the fractal structure
of space-time trajectories seen in numerical simulations, such as
those shown in Figure~\ref{fig:traj}. The dynamical exponent $z$ is
proportional to $1/T$, indicating a rapid slowing-down of the dynamics
near the critical point. As one observes the system on progressively
larger length and time scales, one sees the emergence of a
characteristic length growing as the double exponential of reciprocal
temperature. This length saturates rapidly, and is eventually
supplanted as the characteristic length of the system by by the
equilibrium domain size, whose properties are controlled by the
high-temperature stable fixed point $c^{\star}=1$
$(\lambda^{\star}=\infty)$.}
\end{figure}

With $a=1$ the RG recursion relation~(\ref{recursion-east-0}) may be
iterated to give
\begin{equation}
\label{recursion-east}
\lambda_{p}=\lambda_0 +n \lambda_0^2+\cal{O}\left(\lambda^3\right).
\end{equation}
where $\lambda_{p}$ is the value of the temperature parameter
following a coarse-graining of the system by a factor $p=2^n$. Since
the bare value of $\lambda_0>0$, we see that (\ref{recursion-east})
describes a system with an unstable zero-temperature critical point
$\lambda^{\star}=0$, and a stable high-temperature fixed point
$\lambda^{\star} \rightarrow \infty$. Now, however, the temperature
parameter $\lambda$ is marginally relevant near the fixed point
$\lambda^{\star}=0$. The RG flow diagram is shown in
Figure~\ref{fig:east-flow}.

To determine the correlation length in the East model we proceed as
follows. From the recursion relation (\ref{recursion-east}) we see
that the correlation length satisfies
\begin{equation}
\label{corr-east}
\xi \left( x+ \frac{\ln p}{\ln2} ~ x^2 \right) = p^{-1}\xi(x).
\end{equation}
For small values of $x=\lambda$, corresponding to low temperatures, we
have no solution $\xi(\lambda)$ of (\ref{corr-east}) to first order in
$\lambda$. Thus near criticality the East model possesses no
characteristic length scale. This is consistent with the nature of the
space-time trajectories seen in numerical simulations, such as that
shown in the right panel of Figure~\ref{fig:traj}. These display a
fractal structure~\cite{Garrahan-Chandler}, and hence possess no
characteristic length.

On sufficiently large length and time scales the system will reach
equilibrium, at which point the heights of bubbles will be determined
by the equilibrium spin distribution. This {\em does} have a
characteristic length. One therefore expects to see, sufficiently far
from criticality, the emergence of a length scale. Below, we show that
Equation~(\ref{corr-east}) indeed admits a growing length in such a
regime. This corresponds to the eventual `blurring out' of the fractal
boundaries of clusters as one observes the system on progressively
larger scales. The emerging length scale corresponds to the spatial
extent of bubble regions.

We can quantify the emergence of this length by considering an
infinitesimal RG transformation. The blocking parameter $p$ is
necessarily an integer, because our model is defined on a lattice. But
we can generalize $p$ by considering an infinitesimal change of scale
according to $p=1+\ell$, where $\ell \ll 1$. By writing $\xi(\lambda+
\ell \lambda^2/\ln 2) \approx \xi(\lambda) + \ell d \xi(\lambda)/d
\ell$ and $\lambda'-\lambda \approx \ell d \lambda/d \ell$ we obtain
the flow equations for the temperature and correlation length:
\begin{equation}
\label{temp}
\frac{d \lambda(\ell)}{d \ell} = \lambda(\ell)^2/\ln 2,
\end{equation}
and
\begin{equation}
\label{corr}
\frac{d \xi(\ell)}{d \ell}= -\xi(\ell).
\end{equation}
The intial data for Equations~(\ref{temp}) and~(\ref{corr}) are
$\lambda(\ell_0) = \lambda_0$ and $\xi(\ell_0)=\xi_0$, respectively,
where the subscript zero denotes an unrenormalized (physically
meaningful) quantity. The parameter $\ell_0$ acts as a short distance
regulator (or ultraviolet cutoff), and should be taken to zero at the
end of the calculation.

One now iterates the RG by integrating (\ref{temp}) until $\lambda(\ell) =
\mathcal{O}(1)$, yielding $\ell-\ell_0 \approx \ln2/\lambda_0$. From
(\ref{corr}) we obtain $\xi(\ell) = \xi_0 e^{-(\ell-\ell_0)}$, and so the
correlation length varies with temperature according to \begin{equation}
\xi_0(\lambda_0) \sim \exp\left(1/(\lambda_0 \ln2 )\right) \sim \exp \left(
e^{1/T}/\ln2\right). \end{equation}  Away from the critical point
$\lambda^{\star}=0$ we therefore see an extremely rapid growth of the dynamical
length scale with temperature. 

This length scale corresponds to the emergence of a characteristic
length $\xi_d$ away from criticality, and not to an equilibrium length
scale $l_{eq}$. The latter may be defined as the reciprocal of the
particle density in the steady state (see below), and scales as
$c^{-1}$. The dynamical length is a nonequilibrium critical quantity,
and will be cut off rapidly as one probes larger length and time
scales.  Thus, in terms of the RG flow diagram,
Figure~\ref{fig:east-flow}, the steady-state behaviour is obtained
near the attractive fixed point $\lambda^{\star} \rightarrow \infty$,
where one probes length and time scales much larger than those on
which critical fluctuations are manifest. The critical behaviour will
be observed on short length and time scales, near the critical fixed
point $\lambda^{\star}=0$.

The characteristic equilibration time follows from the relation $\tau
\sim l^{z}$, where $l$ is a typical length scale. We have $z=(T \ln
2)^{-1}$. Taking the equilibrium domain length $l_{eq} \sim
\lambda^{-1}$, we find the equilibration time scale $\tau_{eq} \sim
c^{-z} \sim \lambda^{\ln \lambda}=\exp\{1/(T^2 \ln 2)\}$. This agrees
with results obtained by other means~\cite{Ritort-Sollich}. We assume
that the dynamical exponent $z$ obtained near criticality holds in the
region of the attractive fixed point.

One may also calculate~\cite{Hooyberghs} the density of particles,
$n=\frac{1}{2} (1+\sigma)$. First note that the number operator
renormalizes differently depending on whether $n$ sits in the left or
right slot of the block: $(n \otimes 1)'= n_{\alpha}$, versus $(1
\otimes n)' =c n_{\alpha}$. Hence we will define our density operator
as $n =\frac{1}{2} (n \otimes 1 + 1 \otimes n)$. The RG recursion
relation for the density then reads
\begin{equation}
\label{east-density}
n_{k}= \frac{1}{2}\left( \frac{1+2 \lambda_k}{1+\lambda_k} \right) n_{k+1}.
\end{equation}

To extract the steady-state density we write (\ref{east-density}) as
$n(\lambda_{k})=a(\lambda_k) n(\lambda_{k+1})$, where $a(x)=(1+2x)/\left[2
(1+x) \right]$. By iterating this equation along the RG flow we get
\begin{equation} \label{ss-densityE} n_s(\lambda_0)=\left[ \prod_{i=0}^\infty
a(\lambda_i) \right] n(\lambda^{\star}_a), \end{equation}  where $n_s$ is the
steady-state density, and $n(\lambda^{\star}_a)$ is the density at the
attractive fixed point $\lambda^{\star}_a = \infty$. Next, define $G_n(\lambda)
\equiv \prod_{i=0}^n a(\lambda_i)$. From Equations~(\ref{recursion-east}) and
(\ref{east-density}) we can write \begin{equation} a(\lambda_k) = \frac{1}{2}
\frac{d \ln \lambda_{k+1}}{d \ln \lambda_k}, \end{equation} as in the FA model.
Hence \begin{equation} G_n(\lambda) = \frac{\lambda_0}{\lambda_0+1} \left( 1+
\cal{O}(\lambda_{n+1}^{-1}) \right). \end{equation}  Taking $n \rightarrow
\infty$ gives $\lambda_{n+1} \rightarrow \infty$, and by noting that
$n(\lambda^{\star}_a)= 1$ we obtain the steady-state density \begin{equation}
n_s(\lambda_0)=\frac{\lambda_0}{\lambda_0+1} =c, \end{equation} as expected. 

The behaviour of the density near criticality ($\lambda^{\star}=0$)
follows from the relation $n(\Delta \lambda)=a(\lambda^{\star})
n(\lambda')$, where $\Delta \lambda \equiv
\lambda-\lambda^{\star}=\lambda$. If we iterate the RG until the
renormalized coupling $\lambda'=\cal{O}(1)$, i.e. $\ell/\ell_0 \sim
e^{\ln 2/\lambda_0}$, we find $n(\lambda_0) \sim e^{-\ln
2/\lambda_0}$. Thus the density vanishes close to the critical point
faster than any power of $T$.

\section{Renormalization of the BCIC}
\label{sec:bcic}

In this section we will apply the RG scheme to the BCIC, a model whose
kinetic constraint interpolates between that of the East and FA
models.  We find that on suitably large length and time scales (or for
suitably low temperatures) the BCIC behaves like the FA model. This
agrees with existing numerical and analytical
results~\cite{Buhot-Garrahan}.

The ground state eigenvectors of the BCIC (\ref{interp}) are the same
as those of the FA model. If we use (\ref{proj-fa}) and (\ref{interp})
we find
\begin{equation}
\label{interp-R}
(\cal{H}^{b})'=\frac{1}{2-c}\left(\begin{array}{cccc}
\frac{1}{2-c}-c&-\tilde{b}c&-bc&0\\
\frac{\tilde{b}}{c-2}+\tilde{b}c&\tilde{b}c&0&0\\
\frac{b}{(c-2)}+bc&0&bc&0\\
0&0&0&0\end{array}\right).
\end{equation}
Equations~(\ref{interp}) and~(\ref{interp-R}) yield the same recursion
relation for the temperature parameter $\lambda$ as in the FA model,
$\lambda_{k+1} = \lambda_k(2+\lambda_k)$. They also yield a recursion
relation for the asymmetery parameter $b$: $b_{k+1}=b_k$. Thus the
asymmetry $b$ is a marginal operator, and does not flow under
renormalization. From the RG relation for $\lambda$, we see that for
any $b \in (0,1)$ the interpolation model falls in the universality
class of the FA model, rather than the East model.

However, we expect the interpolation model for small values of $b$ to
display a crossover from East-like to FA-like
behaviour~\cite{Buhot-Garrahan}. This suggests that by projecting
$\cal{H}^{b}$ onto a subspace spanned by only the ground states of
(\ref{interp}) we have omitted this crossover behaviour. We can
recover it in the following way.

First, we note that the difference between the East and FA models
manifests itself in the treatment of the states $\ud$ and $\du$ during
embedding. In the East model the latter is completely suppressed (see
Equation (\ref{proj})); in the FA model, both are treated on equal
footing [Eq.~(\ref{proj-fa})]. By restricting our RG scheme to a
subspace of the ground states of Equation (\ref{interp}), we are
unable to construct an embedding operator that treats $\du$ and $\ud$
asymmetrically [cf. $R_2$, Equation (\ref{proj})].

To remedy this, we now include the first excited right eigenvector of
(\ref{interp}) in our embedding operator. We will call this
eigenvector $e$. This is akin to calculating higher-order ``loop''
diagrams to check if $b$, ostensibly a marginal operator, is relevant
at second order. $e$ has eigenvalue $2(1-b) bc + \cal{O}(c^2)$, and is
therefore a ``gapless excitation'' in the East model limit, $b
\rightarrow 0$. Note that $e = (e_1,e_2,1,0)^{T}$, where the $e_i$ are
functions of $c$ and $b$. For small $c$ we have $e \approx \left[
(2b-1)c,-1+(1-2b)c,1,0 \right]$.

Let us now construct a new embedding operator,
\begin{equation}
\label{proj-interp}
\tilde{R}_2=\frac{1}{2-c}\left(\begin{array}{cc}
 c+\alpha e_1&0\\
1-c+\alpha e_2&0\\
1-c+\alpha &0\\
0&2-c\end{array}\right),
\end{equation}
and demand that in the limits $b \rightarrow 0$ and $b \rightarrow
\frac{1}{2}$ we recover the respective embedding operators for the
East and FA models, namely $R_2$ and $T_2$. This is achieved by
setting $\alpha=-(1-2 b)(1-c)$. We note that $R_1
\tilde{R}_2=\mathbf{1}$.

Our renormalization prescription is now $(\cal{H}^b)'=R_1 \cal{H}^b
\tilde{R}_2$. We derive recursion relations for the parameters $c$ and
$b$ in a similar way to before: we define the unrenormalized
temperature parameter $\lambda$ as the ratio
\begin{equation}
\lambda \equiv -\frac{(1,2)+(1,3)}{(1,1)}=\frac{c}{1-c},
\end{equation}
where $(i,j)$ is element $(i,j)$ of the matrix $\cal{H}_b \equiv
\cal{L}$, Equation (\ref{interp}). We define the renormalized
parameter $\lambda'$ by the ratio of the corresponding elements of the
renormalized matrix $(\cal{H}^b)'$. This gives us the recursion
relation $\lambda_{k+1}=f(\lambda_k,\mu_k)$. The parameter $\mu$ is
the scaled asymmetry parameter, whose unrenormalized value we define
as
\begin{equation}
\mu \equiv \frac{(1,3)}{(1,2)}=\frac{b}{1-b}.
\end{equation}
The elements $(i,j)$ again refer to Equation (\ref{interp}). We write
the recursion relation for $\mu$, obtained from the elements of
$(\cal{H}^b)'$, as $\mu_{k+1}=g(\lambda_k,\mu_k)$.

\begin{figure}
\psfig{file=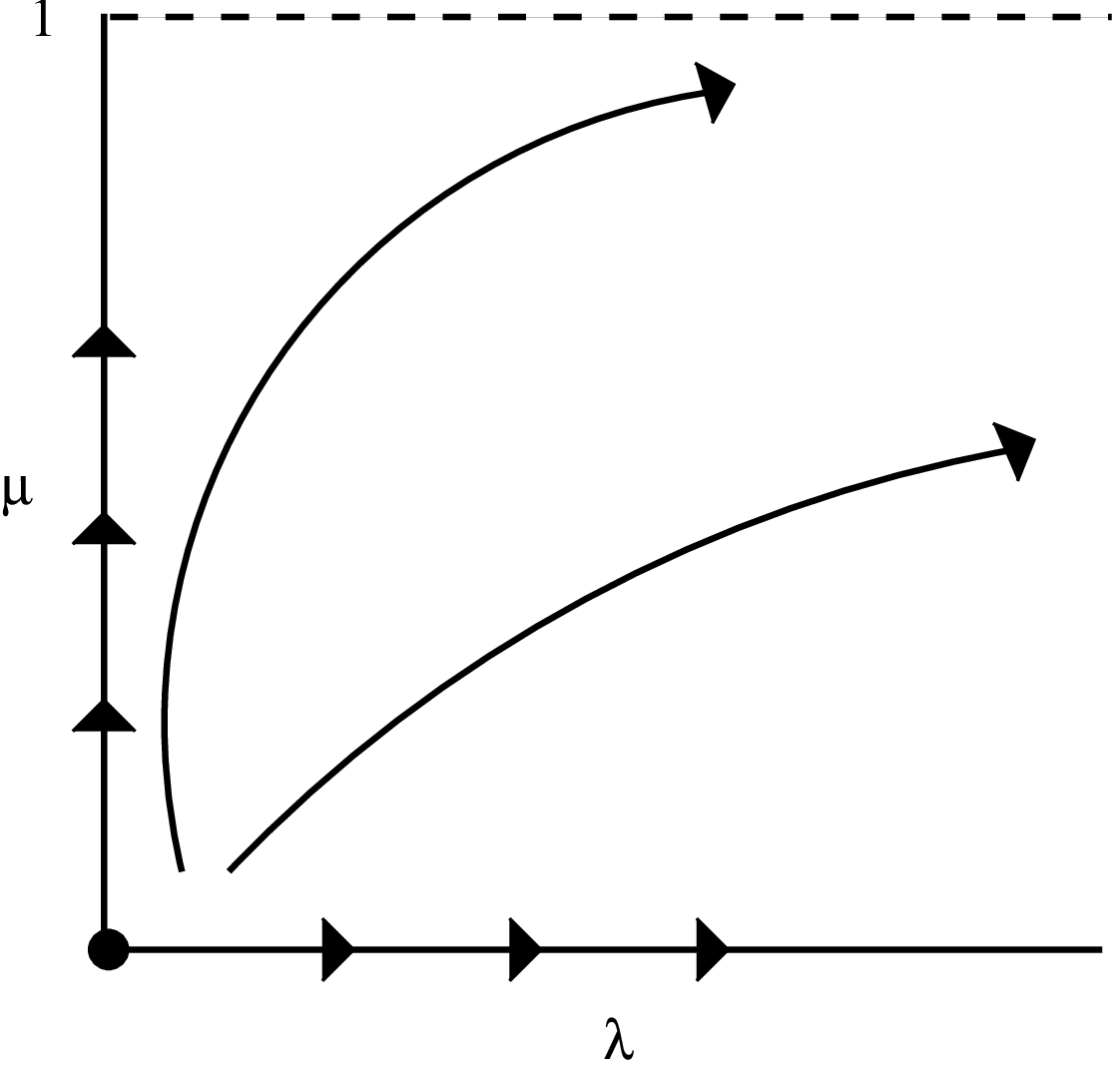,width=4cm,angle=270}
\caption{\label{fig:bcic-flow} RG flow diagram for the BCIC, in
$(\lambda,\mu)$ space. $\mu=0$ (resp. 1) corresponds to East
(resp. FA) model behaviour. The critical fixed point $(0,0)$ is
unstable; the attractive fixed point $(\infty, 1)$ corresponds to the
high-temperature fixed point of the FA model.}
\end{figure}

The behaviour of the functions $f$ and $g$ thus determine the
crossover properties of our model. We find that $\lambda$ has an
unstable zero-temperature fixed point $\lambda^{\star}=0$, and an
attractive high-temperature fixed point $\lambda^{\star} \rightarrow
\infty$. The asymmetry $\mu$ has an unstable maximal-aysymmetry fixed
point $\mu^{\star}=0$, corresponding to the East model, and an
attractive symmetric fixed point $\mu^{\star}=1$, corresponding to the
FA model. Thus any BCIC with less than maximal asymmetry will behave
at long length and time scales like the FA
model. Figure~\ref{fig:bcic-flow} shows the qualitative RG flow of the
BCIC.

For the case of $p=2$ we find that
\begin{equation}
\label{recursion-temp}
\lambda_{k+1}= \left\{ \begin{array}{lr} \lambda_k(1+\lambda_k) +
f_1(\lambda_k) \mu_k +\cal{O}(\mu_k^2), & \mu_k \approx 0; \\
\lambda_k(2+\lambda_k)+f_2(\lambda_k) \tilde{\mu}_k^2
+\cal{O}\left(\tilde{\mu}_k^3\right),& \mu_k \approx 1,
\end{array} \right.
\end{equation}
where $f_1(x) \equiv (2+9x+11x^2+6x^3+x^4)(2+3x+x^2)^{-1}$, $f_2(x)
\equiv x^2(x-1)(2+x)/[8(1+x)]$, and $\tilde{\mu}_k \equiv
1-\mu_k$. Equations (\ref{recursion-temp}) thus reproduce the
recursion relations for $\lambda$ in the East model and FA model
limits, respectively Equations (\ref{recursion-east-0}) and
(\ref{recursion-fa}). The asymmetry parameter $\mu$ is a relevant
perturbation, whose flow is governed by
\begin{equation}
\label{assym}
\mu_{k+1}=\left\{ \begin{array}{lr} \left(1+\frac{1}{\lambda_k}
\right) \mu_k-f_3(\lambda_k) \mu_k^2+\cal{O}(\mu_k^3), & \mu_k \approx
0;\\ 1-\frac{\lambda_k}{1+\lambda_k}
\tilde{\mu}_k+\cal{O}\left(\tilde{\mu}_k^2\right), & \mu_k \approx 1,
\end{array} \right. 
\end{equation}
where $f_3(x) \equiv (2+4x+5x^2+3x^3)/[x^2(1+x)(2+x)]$. 

 We can deduce the flow of the BCIC away from maximal asymmetry, $\mu
 =0$, by studying Equations (\ref{recursion-temp}) and (\ref{assym})
 in the regime $\mu \ll \lambda \ll 1$. Writing $\beta_{\lambda}\equiv
 \lambda'-\lambda \approx \ell d\lambda/d\ell$, and a similar relation
 for $\mu$, we obtain
\begin{eqnarray}
\label{first}
\beta_{\lambda}&=&\ell \lambda^2+\mu+\cal{O}(\lambda
\mu)+\cal{O}(\mu^2/\lambda); \\
\label{second}
\beta_{\mu}&=&\frac{\mu}{\lambda}-\frac{\mu^2}{\lambda^2} +
\cal{O}(\mu^3/\lambda^3)+\cal{O}(\mu^2/\lambda).
\end{eqnarray}
Equations~(\ref{first}) and~(\ref{second}) may be solved in terms of
the exponential integral function $Ei\left(\lambda^{-1} \right)$,
although the physical interpretation of this solution is not
obvious. We can more clearly determine the essence of the crossover as
follows.

The temperature parameter $\lambda$ has RG eigenvalue 0 (East model)
or 1 (FA model). It therefore grows much less rapidly than the
asymmetry parameter $\mu$, which has (initial) eigenvalue
$\lambda_0^{-1} \gg 1$. Hence from (\ref{assym}) we have
$\mu'=2^{y_{\mu}} \mu \approx \lambda_0^{-1} \mu$, giving the RG
eigenvalue for the asymmetry parameter as $y_{\mu} \approx \left(T
\ln2\right)^{-1}$. Let us now write a standard RG scaling form for the
particle density,
\begin{equation}
\label{rg-eqn-density}
n_R(\lambda',\mu')=p \, n\left(p^{y_{\lambda}} \lambda,p^{y_{\mu}}
\mu,p^{-1} \xi,p^{-z} t \right)
\end{equation}
To derive a crossover temperature, we iterate the RG until
$p^{y_{\lambda}} = \cal{O}(1)$. The $\mu-$dependent scaling
combination is then $\lambda^{-y_{\mu}/y_{\lambda}}$. When this
becomes large, i.e. $\cal{O}(1)$, one would expect the BCIC to behave
like the FA model. Taking for simplicity $y_{\lambda}=1$, we find a
crossover temperature $T_{xo} \sim \left(-\ln \mu
\right)^{-1/2}$. This scaling agrees with that obtained by equating
the relaxation timescale for the $\mu$-suppressed symmetric process,
$\tau_S \sim \left( \mu \lambda \right)^{-1}$ with that for the
asymmetric process, $\tau_A \sim \exp \left(1/T^2 \ln
2\right)$~\cite{Buhot-Garrahan,Garrahan-Chandler}.

We can extract crossover time- and length-scales from
Equation~(\ref{rg-eqn-density}) by iterating the RG until,
respectively, $p^{-1} \xi =\cal{O}(1)$ and $p^{-z} t
=\cal{O}(1)$. These give $\xi_{xo} \sim \mu^{-T \ln2}$ and $t_{xo}
\sim \mu^{-2 T \ln2}$.

The real-space RG therefore confirms that for anything less than
maximal asymmetry, the BCIC will on long length and time scales
display FA-like, as opposed to East-like
behaviour~\cite{Buhot-Garrahan}.

\section{Conclusions}
\label{sec:conclu}

We have used the simple real-space RG scheme of Refs.~\cite{Stella,Hooyberghs}
to derive the zero-temperature critical behaviour of the FA, East and BCIC
models.  Our findings agree with known
results~\cite{Sollich-Evans,Chung-et-al,Buhot-Garrahan,Garrahan-Chandler,Ritort-Sollich}, but offer a different and unified approach to these systems.  We are
also aware of alternative real-space RG studies of KCICs \cite{Hans,Robin}. 

The real-space RG scheme used in this paper is sufficiently flexible
to be extended to more complicated models.  An interesting possibility
would be to use this scheme to study a recently-introduced model of
the reentrant glass transition in colloids~\cite{Phill}, which
combines dynamical constraints with static interactions.

\section{Acknowledgements}
We are grateful to Hans Andersen, Robert Jack, and Robin Stinchcombe
for discussions.  We acknowledge financial support from EPSRC Grants
No.\ GR/R83712/01 and GR/S54074/01, and University of Nottingham Grant
No. FEF 3024.


\begin{thebibliography}{99}

\bibitem{Fredrickson-Andersen} G.H. Fredrickson and H.C. Andersen,
Phys. Rev. Lett. {\bf 53}, 1244 (1984); J. Chem. Phys. {\bf 83}, 5822
(1985).

\bibitem{Palmer-et-al} R.G Palmer, D.L. Stein, E. Abrahams and
P.W. Anderson, Phys. Rev. Lett. {\bf 53}, 958 (1984).

\bibitem{Jackle-Eisinger} J. J\"ackle and S.  Eisinger, Z. Phys. {\bf
  B84}, 115 (1991).

\bibitem{Kob-Andersen} W. Kob and H.C. Andersen, Phys. Rev. E {\bf
48}, 4364 (1993).

\bibitem{Whitelam-Garrahan} S. Whitelam and J.P. Garrahan,
J. Phys. Chem. B {\bf 108}, 6611 (2004).

\bibitem{Schulz-Trimper} M. Schulz and S. Trimper, J. Stat. Phys. {\bf
  94} 173 (1999).

\bibitem{Sollich-Evans} P. Sollich and M. R. Evans, Phys. Rev. Lett. {\bf
83}, 3238 (1999); Phys. Rev. E 68, 031504 (2003).

\bibitem{Crisanti-et-al} A. Crisanti, F. Ritort, A. Rocco and
M. Sellitto, J. Chem. Phys. {\bf 113}, 10615 (2001).

\bibitem{Chung-et-al} F. Chung, P. Diaconis and R. Graham,
Adv. App. Math., {\bf 27}, 192, (2001).

\bibitem{Toninelli-et-al} C. Toninelli, G. Biroli and D.S. Fisher,
Phys. Rev. Lett. {\bf 92}, 185504 (2004).

\bibitem{Garrahan-Chandler} J.P. Garrahan and D. Chandler,
Phys. Rev. Lett. {\bf 89}, 035704 (2002); Proc. Natl. Acad. Sci. USA
{\bf 100}, 9710 (2003).

\bibitem{Berthier-Garrahan} L. Berthier and J.P. Garrahan,
J. Chem. Phys. {\bf 119}, 4367 (2003);  Phys. Rev. E 68, 041201 (2003).

\bibitem{Ritort-Sollich} F. Ritort and P. Sollich, Adv. in Phys. {\bf
52}, 219 (2003).

\bibitem{Hooyberghs} J. Hooyberghs and C. Vanderzande, J. Phys. A.,
{\bf 33}, 907, (2000); Phys. Rev. E, {\bf 63}, 041109, (2001).

\bibitem{Whitelam-et-al} S. Whitelam, L. Berthier and J.P. Garrahan,
Phys. Rev. Lett. {\bf 92}, 185705 (2004).

\bibitem{Carlon-et-al} E. Carlon, M. Henkel and U. Schollwoeck,
Eur. Phys. J. B12, 99 (1999).

\bibitem{Tauber} U.C. T\"{a}uber, Adv. in Solid State Physics,
B. Kramer (Ed.), Vol 43 (Springer-Verlag Berlin), {\bf 659}, 675
(2003), \texttt{cond-mat/0304065}.

\bibitem{Cardy-rev} J.L. Cardy, Renormalisation group approach to
reaction-diffusion problems, {\texttt cond-mat/9607163} (1996).

\bibitem{Siggia} E.D. Siggia, Phys. Rev. B, {\bf 16}, 2319 (1977).

\bibitem{Buhot-Garrahan} A. Buhot and J.P. Garrahan, Phys. Rev. E 64,
21505 (2001).

\bibitem{Hinrichsen} H. Hinrichsen, Adv. Phys. {\bf 49}, 815 (2000).

\bibitem{Stella} A. Stella, C. Vanderzande and R. Dekeyser,
Phys. Rev. B, {\bf 27}, 1812 (1983).

\bibitem{Cardy} J.L. Cardy, {\em Scaling and Renormalization in
Statistical Physics} (Cambridge University Press, Cambridge, 1996).

\bibitem{Kadanoff} L.P. Kadanoff, {\em Statistical Physics: Statics,
Dynamics and Renormalization} (World Scientific, Singapore, 2000).

\bibitem{Aldous-Diaconis} D. Aldous and P. Diaconis,
J. Stat. Phys. {\bf 107}, 945 (2002).

\bibitem{Hans} H.C. Andersen, private communication.

\bibitem{Robin} R. Stinchcombe, private communication.

\bibitem{Phill} P.L. Geissler and D.R. Reichman, e-print
\texttt{cond-mat/0402673}.

\end{thebibliography}
\end{document}